\documentclass{article}
\usepackage[utf8x]{inputenc}
\usepackage{spconf,amsmath,graphicx}

\graphicspath{{figures/}}

\usepackage{cite}
\usepackage{amsmath,amssymb,amsfonts}
\usepackage{algorithmic}
\usepackage{graphicx}
\usepackage{multirow}
\usepackage{tikz}
\usepackage{hyperref}
\usepackage{xcolor}
\usepackage{lipsum}
\usepackage{caption}
\usepackage{siunitx}
\usepackage{commath}
\usepackage{subfig}
\usepackage{pgfplots}
\pgfplotsset{compat=1.17}
\usepackage{mathrsfs}
\usetikzlibrary{arrows}
\usetikzlibrary{positioning}
\usetikzlibrary{3d}
\usepackage{comment}
\usepackage{adjustbox}
\usepackage{textcomp}
\usepackage{xcolor}
\def\BibTeX{{\rm B\kern-.05em{\sc i\kern-.025em b}\kern-.08em
    T\kern-.1667em\lower.7ex\hbox{E}\kern-.125emX}}
 
\begin{document}

\definecolor{col_w}{rgb}{0.870588,0.796078,0.776470}
\definecolor{col_wn}{rgb}{0.996078,0.847059,0.364706}
\definecolor{col_out}{rgb}{1,0.5,0}
\definecolor{col_bc}{rgb}{0,0.5,1}
\definecolor{col_conv}{rgb}{0,1,1}
\definecolor{col_out}{rgb}{1,0.5,0}

\title{The Practice of Averaging Rate-Distortion Curves over Testsets to Compare Learned Video Codecs Can Cause Misleading Conclusions}

\name{M. Akin Yilmaz$^{\star}$ \qquad Onur Keleş$^{\star \dagger}$ \qquad A. Murat Tekalp$^{\dagger}$
\thanks{A. M. Tekalp acknowledges support from the Turkish Academy of Sciences (TUBA).}}

\address{$^{\star}$ Codeway AI Research \\
      $^{\dagger}$ Dept. of Electrical \& Electronics Engineering, Koç University, İstanbul, Türkiye 
}

\maketitle
\begin{abstract}
This paper aims to demonstrate how the prevalent practice in the learned video compression community of averaging rate-distortion (RD) curves across a test video set can lead to misleading conclusions in evaluating codec performance. Through analytical analysis of a simple case and experimental results with two recent learned video codecs, we show how averaged RD curves can mislead comparative evaluation of different codecs, particularly when videos in a dataset have varying characteristics and operating ranges. We illustrate how a single video with distinct RD characteristics from the~rest of the test set can disproportionately influence the~average RD curve, potentially overshadowing a codec's superior performance across most individual sequences. Using two recent learned video codecs on the UVG dataset as a case study, we demonstrate computing performance metrics, such as the~BD rate, from the average RD curve suggests conclusions that contradict those reached from calculating the~average of per-sequence metrics. Hence, we argue that the learned video compression community should also report per-sequence RD curves and performance metrics for a test set should be computed from the average of per-sequence metrics, similar to the established practice in traditional video coding, to ensure fair and accurate codec comparisons.

\end{abstract}

\section{Introduction}
Comparing performances of different video codecs is a critical yet challenging task in video compression research. While measuring performance through rate-distortion (RD) curves provides insights into codec behavior, the methodology of presenting and analyzing these curves can significantly impact conclusions. The Bjøntegaard Delta (BD) \cite{bdrate} measure, which calculates the average difference between RD curves, where distortion is typically quantified by peak signal-to-noise ratio (PSNR) \cite{psnr_comp}, is commonly used to quantify these comparisons.

The classic video compression community evaluate codecs by computing RD curves and performance metrics, such as the BD rate, per-video, then averaging these metrics across a dataset. In contrast, the learned video compression community commonly present an RD curve averaged across all videos in a dataset, and then compute a single BD rate from this average RD curve. This latter practice, while appearing to be a convenient approach, can lead to misleading conclusions about codec performance.
When RD curves are averaged across videos with varying characteristics, even a single video can dramatically skew the average curve. For instance, a video with distinct RD characteristics can disproportionately influence the average, potentially masking a codec's superior performance across most sequences. This problem becomes particularly severe when different codecs operate at varying bitrate ranges - a common scenario in learned compression where models may adapt differently to content characteristics. 
To ensure fair and accurate codec assessment, we advocate for following the traditional video coding community's approach: computing RD curves and performance metrics for each video individually before averaging. 

In the following, Section~\ref{sec:related} reviews related work and discusses contributions of this paper, Section~\ref{sec:math} presents an analytical analysis of a simple case demonstrating how averaged RD curves can lead to misleading conclusions, and Section~\ref{sec:results} offers experimental evidence comparing two recent learned video codecs, revealing how computing BD rate from averaged RD curves leads to a different conclusion compared to averaging per-sequence BD-rates. 
By highlighting the pitfalls of averaging RD curves and advocating for evaluation of average BD-rates as average of BD-rates calculated from per-sequence RD curves, we hope to influence the learned video compression community to adopt the more accurate and reliable assessment methodology that has been a common practice in the classic video coding community.

\section{Related work and Contributions}
\label{sec:related}
Traditional video codecs, such as those described by the High Efficiency Video Coding (HEVC) \cite{hevc} and the Versatile Video Coding (VVC) \cite{vvc} standards, employ rigorous evaluation methodologies where codec performance is assessed on a per-sequence basis. These conventional codecs are compared using consistent metrics computed individually for each sequence, ensuring fair assessment across different content types. In contrast, the learned video compression community has widely adopted a different evaluation methodology. Despite significant advances in compression techniques through deep learning approaches \cite{agustsson_scale, rlvc, elfvc, FVC2022,ladune2021conditional,li2021deep, li2022hybrid, sheng2022temporal, li2023neural}, the evaluation practices have diverged from traditional practices. 

Examination of recent literature reveals the prevalence of averaging RD curves across entire datasets before performance analysis. For instance, Chen and Peng \cite{canfvcpp} explicitly stated "The average bpp and PSNR-RGB are computed across the entire dataset, treating it as a single input sequence." This practice is evident in numerous influential works, including scale-space flow methods \cite{agustsson_scale}, hierarchical quality enhancement approaches \cite{hlvc}, and neural representation-based compression \cite{hinerv, chen2021nerv}.
The issue extends beyond individual papers to affect the entire field. Recent works in learned bi-directional coding \cite{ours_icip20, hlvc, lhbdc, flexrate, ours_icip23} predominantly rely on averaged RD curves for performance assessment. This practice has become so common that even some benchmark platforms \cite{pytorchvideocompression} primarily present averaged curves rather than per-sequence analyses.

The main contribution of this paper is to demonstrate why the practice of averaging RD curves can lead to misleading conclusions about codec performance. Specifically: \vspace{-4pt}
\begin{itemize}
\item We provide an analytical analysis in Section~\ref{sec:math} showing how averaged RD curves can produce inconsistent results, particularly when codecs operate in different bitrate ranges - a common case in learned compression. \vspace{-18pt}
\item Through experiments in Section~\ref{sec:results}, we demonstrate how the practice of averaging RD curves can overshadow performance differences between codecs, using two recent learned compression methods as a case study. 
\end{itemize}
\vspace{-4pt}

We would like to emphasize that while the per-sequence computation of RD curves and BD rates is an established approach in the traditional video coding community, {\it there is no published literature that shows the pitfalls of computing a single BD rate from the average RD curve}, which is the de facto approach in the learned video compression community.

\section{Why Averaging RD Curves Misleads Codec Assessment: \\ Analysis for a Simple Case}
\label{sec:math}

We provide analysis for the simple scenario with two hypothetical codecs, codec-1 and codec-2, whose RD curves are linear over the bitrate range of interest for two videos, video-1 and video-2. 
We assume that the bitrate ranges for both codecs to encode video-1 fully overlap, and both codecs give exactly the~same RD points for video-1 with rate $R_1^i$ and PSNR $P_1^i$ for $i=\{1,2,\dots,N\}$.  For video-2, let the~RD points for codec-1 be $(R_2^{i}, P_2^{i})$, and RD points of codec-2 be $(R_2^{i+1}, P_2^{i+1})$ for $i=\{1,2,\dots,N\}$. Here, we assume that the rate points for codec-2 are shifted up by one rate point compared to codec-1.
Suppose that the following relationships hold:
\begin{eqnarray}
    R_1^{i+1} - R_1^{i} = \Delta B_1 > 0 &\text{for} &i=\{1,2,\dots,N-1\} \nonumber \\ 
    P_1^{i+1} - P_1^{i} = \Delta P_1 > 0 &\text{for} &i=\{1,2,\dots,N-1\} \nonumber \\
    R_2^{i+1} - R_2^{i} = \Delta B_2 > 0 &\text{for} &i=\{1,2,\dots,N\} \nonumber \\
    P_2^{i+1} - P_2^{i} = \Delta P_2 > 0 &\text{for} &i=\{1,2,\dots,N\} \nonumber 
\end{eqnarray}
which ensure that the RD curves for both codecs on each video are linear. 

The BD-rate for the first video is $0$ since RD curves are exactly the same, and for video-2, $N-1$ points overlap making the RD curves on this range the same, thus, resulting in $0$ BD-rate on this range.
Although the two codecs can be assessed as equivalent in this particular case according to the BD-rates on individual videos, the result can be different when the average RD curve is considered. Since each video has a linear RD curve, the average RD curve is also linear, which can be represented by the equation $P = m\cdot R + b$.

For the first codec, the first two points on the average RD curve are
\begin{equation}
    \left\{\left(\dfrac{R_1^{1}+R_2^{1}}{2}, \dfrac{P_1^{1}+P_2^{1}}{2}\right)\right\}\nonumber
\end{equation}
and
\begin{equation}
    \left\{\left(\dfrac{R_1^{1}+R_2^{1}+\Delta B_1+\Delta B_2}{2}, \dfrac{P_1^{1}+P_2^{1}+\Delta P_1+\Delta P_2}{2}\right)\right\}.\nonumber
\end{equation}

Similarly, for the second codec, the first two points on the average RD curve are
\begin{equation}
    \left\{\left(\dfrac{R_1^{1}+R_2^{1}+\Delta B_2}{2}, \dfrac{P_1^{1}+P_2^{1}+\Delta P_2}{2} \right)\right\}\nonumber
\end{equation}
and
\begin{equation}
    \left\{\left(\dfrac{R_1^{1}+R_2^{1}+\Delta B_1+2\Delta B_2}{2}, \dfrac{P_1^{1}+P_2^{1}+\Delta P_1+2\Delta P_2}{2}\right)\right\}.\nonumber
\end{equation}

Both curves have slope $m = (\Delta P_1+\Delta P_2)/(\Delta B_1+\Delta B_2)$. For these curves to overlap and show equivalent performance, their y-intercepts must be equal. After substituting these points into $P = mR + b$ and solving for the intercepts, we find that they are equal only when:
\begin{equation}
\label{eq:lin_cond}
\Delta P_2 \cdot \Delta B_1 = \Delta P_1\cdot \Delta B_2
\end{equation}
Even in this simple case, if this condition is not satisfied, then the average RD curves will indicate different performance levels, contradicting the per-sequence evaluation results where the codecs perform identically.



\begin{figure}[t]
\centering
    \includegraphics[scale=0.245]{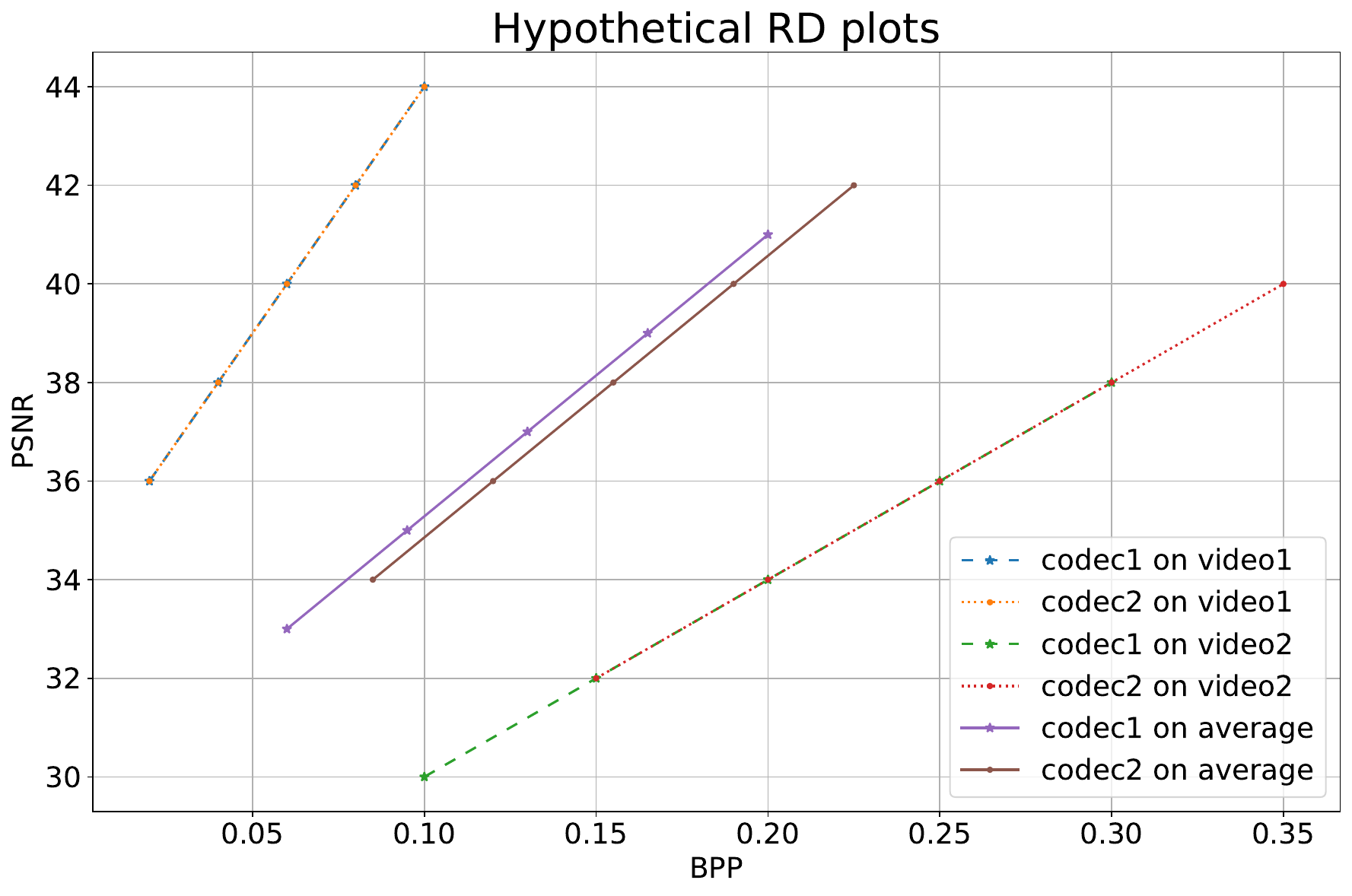} \vspace{-6pt} \\
\caption{Illustration of averaging RD curves in the case of two hypothetical codecs with linear RD curves, where the bitrate range of two codecs to encode video-2 do not fully overlap.}
\label{fig:imaginary}
\end{figure}

We illustrate the simplistic scenario in the case of hypothetical codecs with linear RD curves in Fig.~\ref{fig:imaginary}. We depict the RD curves for video-1 on the left, video-2 on the right, and the average RD curves in the middle. These plots illustrate how, even when the codecs exhibit equal performance for each individual video, a violation of the condition stated in Eq.~\eqref{eq:lin_cond} can lead to a misleading conclusion that the average RD curve of one codec to outperform the other.

\begin{figure*}[t]
\centering
\subfloat      {
\includegraphics[width=.248\linewidth]{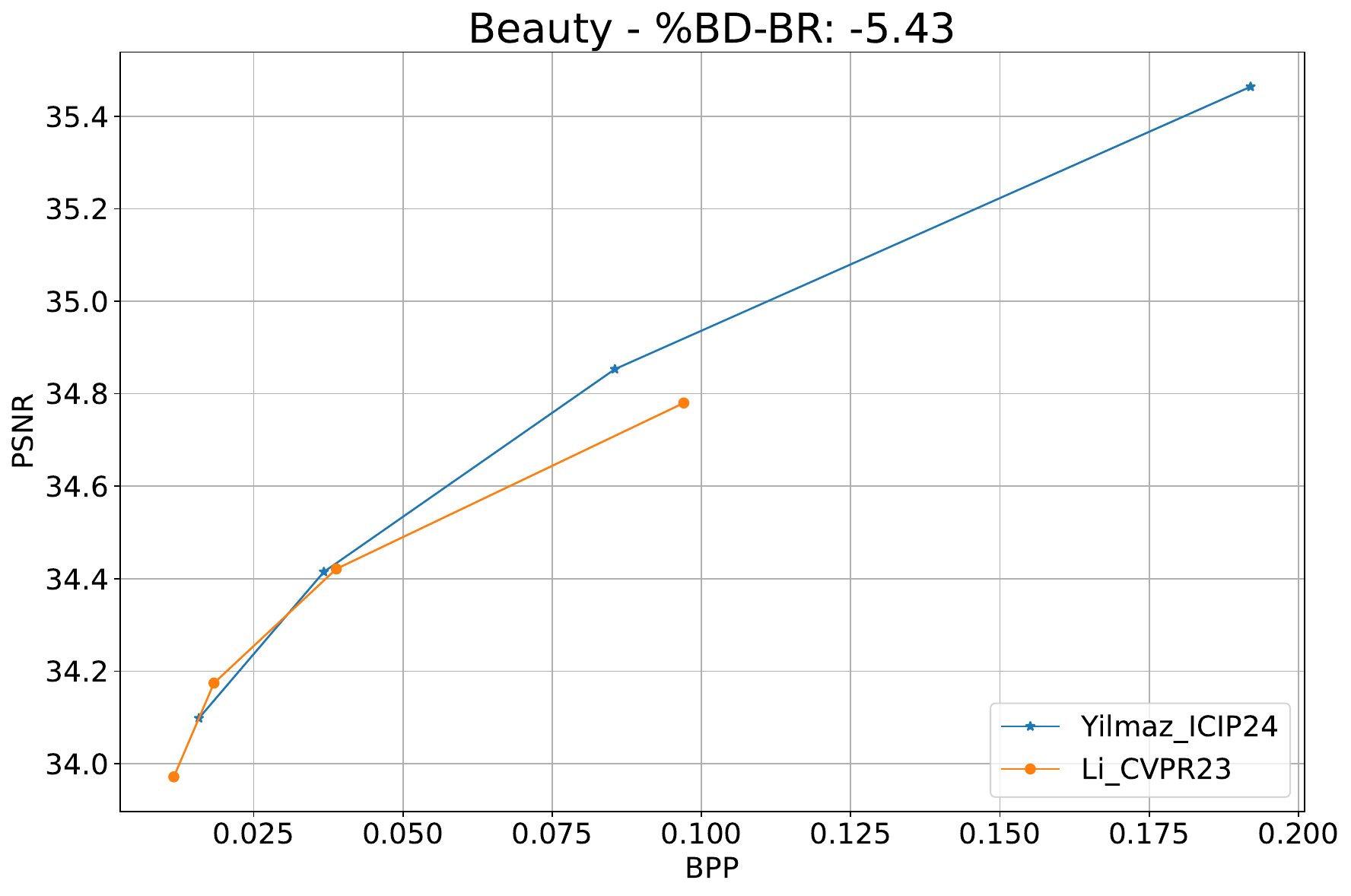}
} \hspace{-8pt}
\subfloat     {
\includegraphics[width=.248\linewidth]{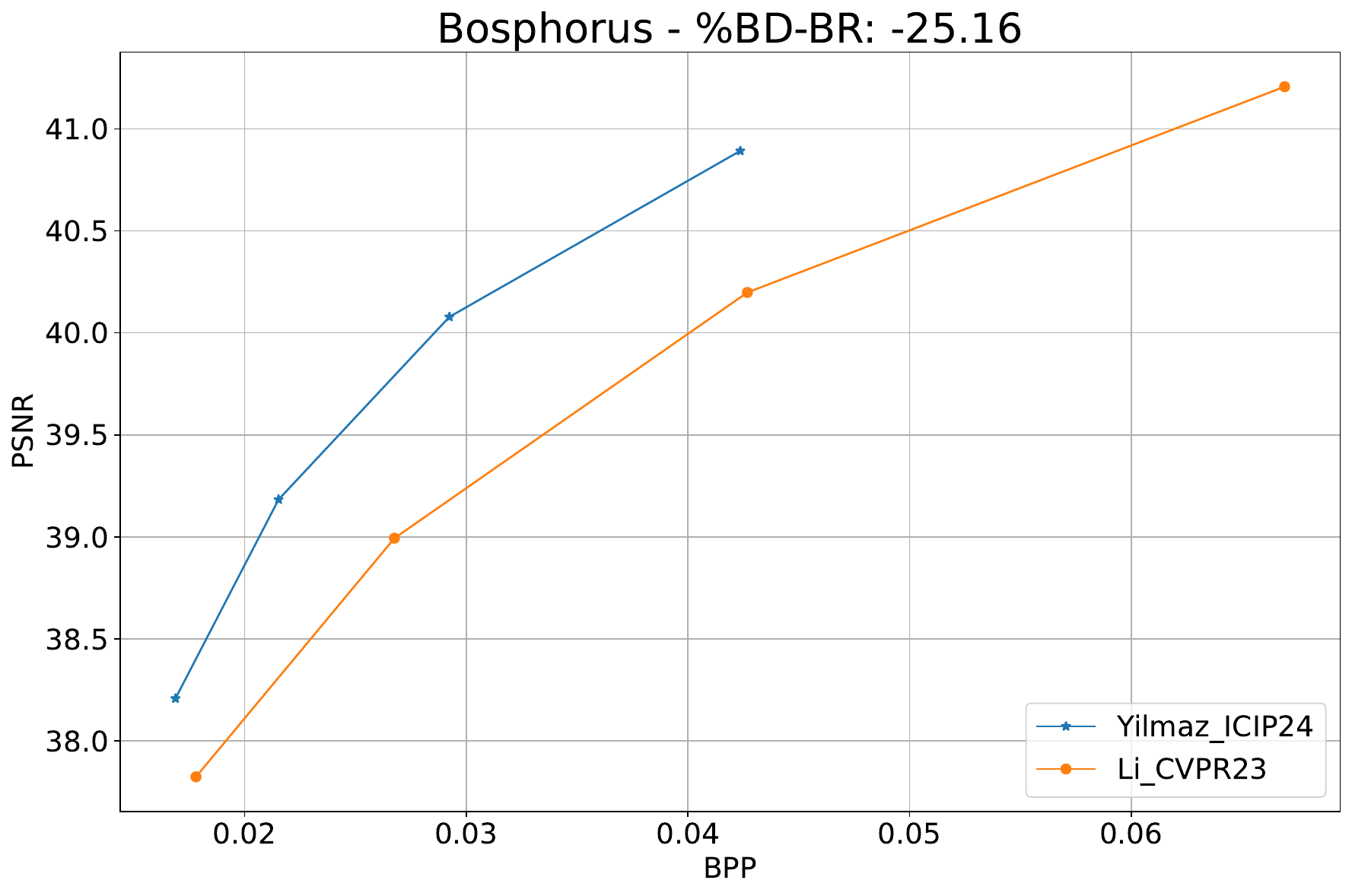}
} \hspace{-8pt}
\subfloat     {
\includegraphics[width=.248\linewidth]{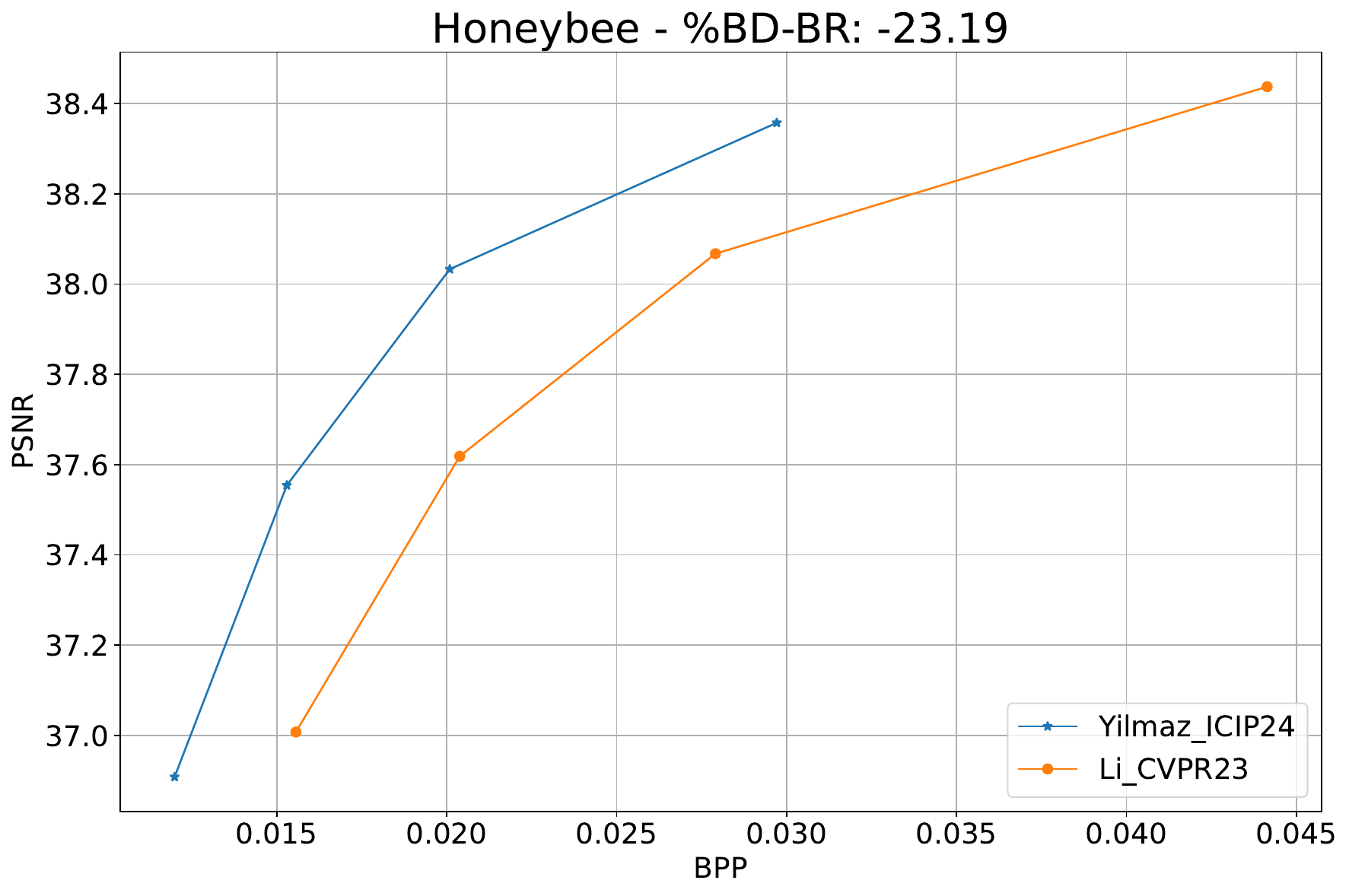}
} \hspace{-8pt} 
\subfloat     {
\includegraphics[width=.248\linewidth]{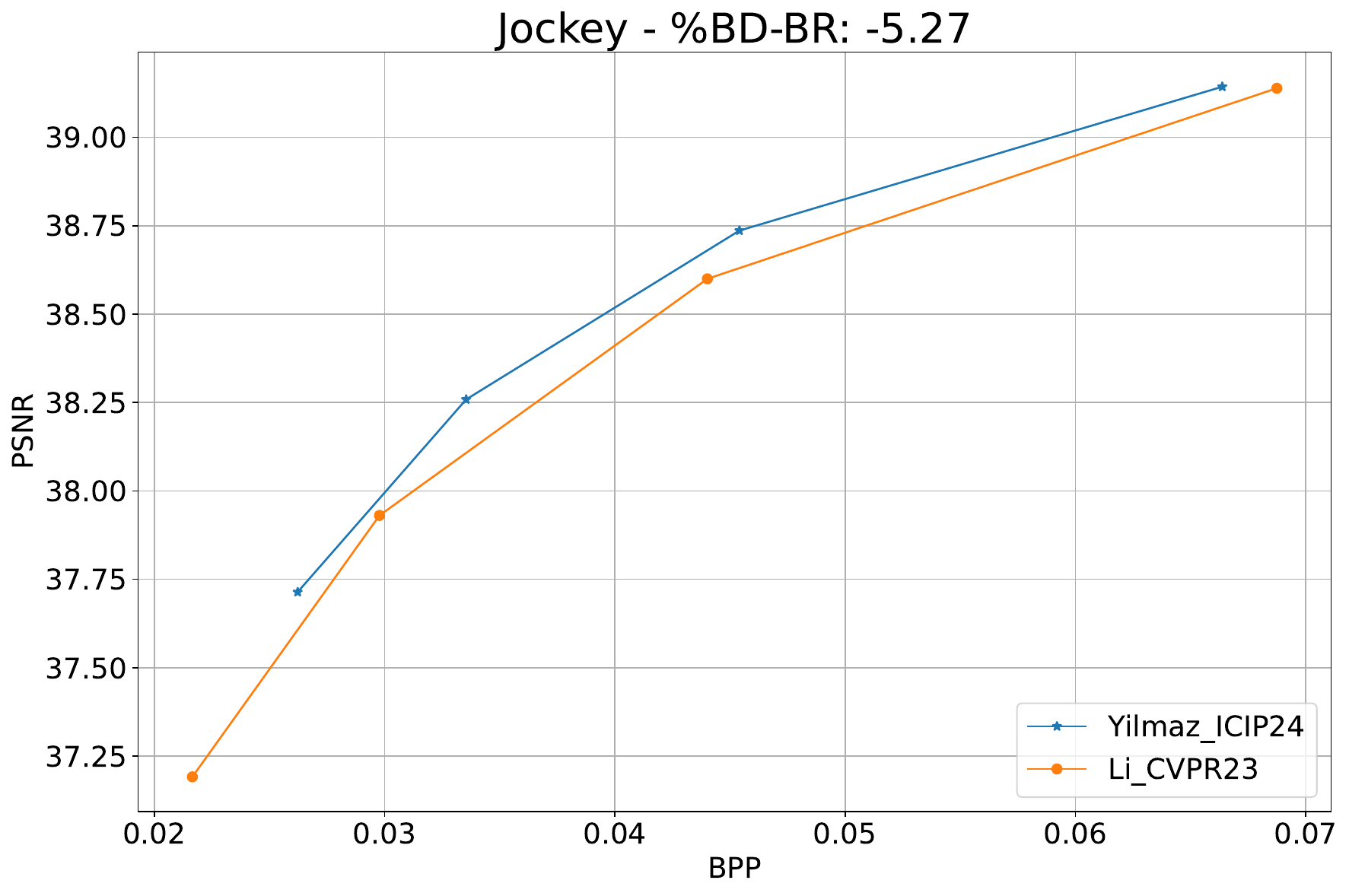}
} \vspace{-8pt} \\
\subfloat     {
\includegraphics[width=.248\linewidth]{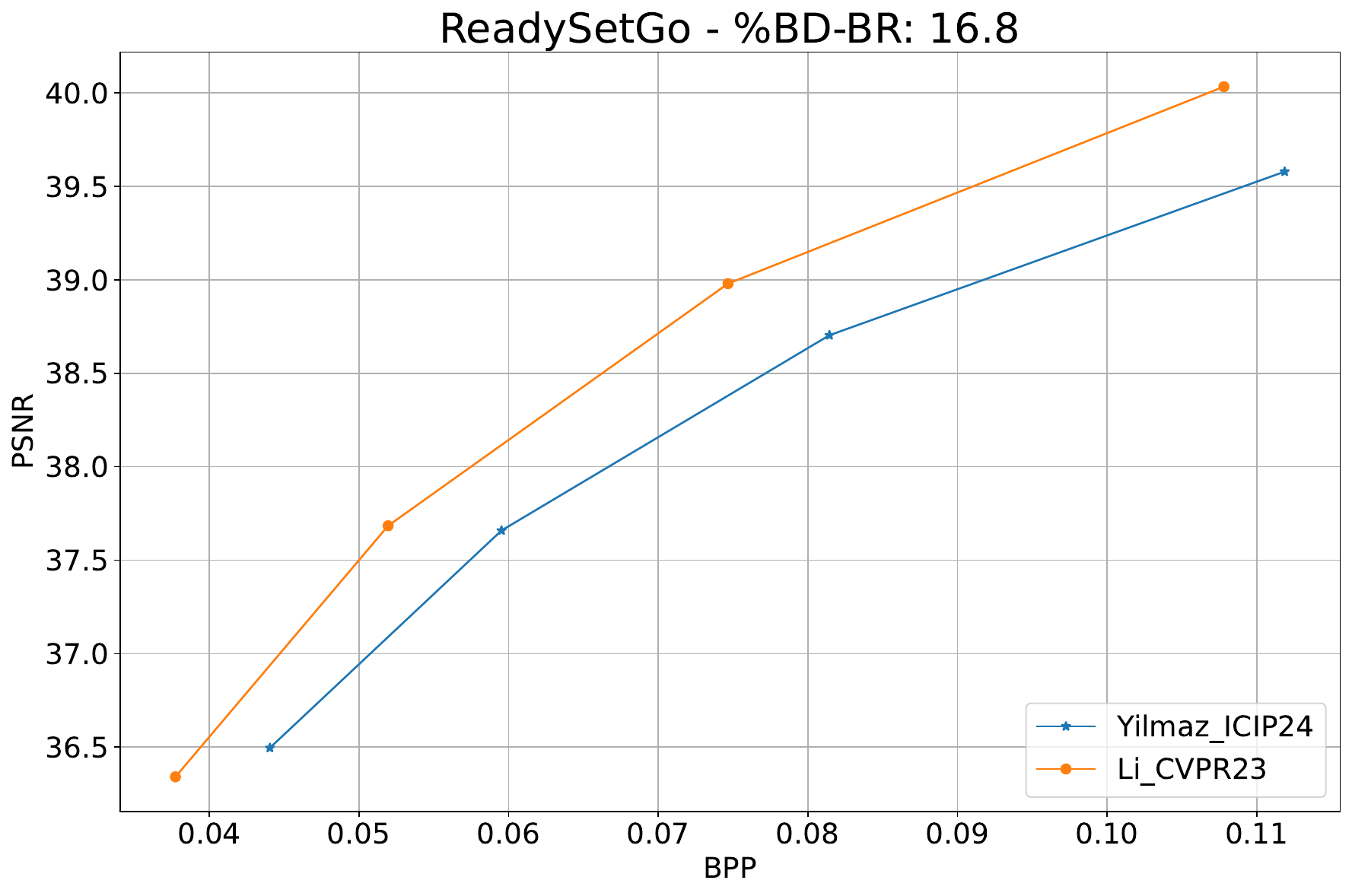}
} \hspace{-8pt}
\subfloat     {
\includegraphics[width=.248\linewidth]{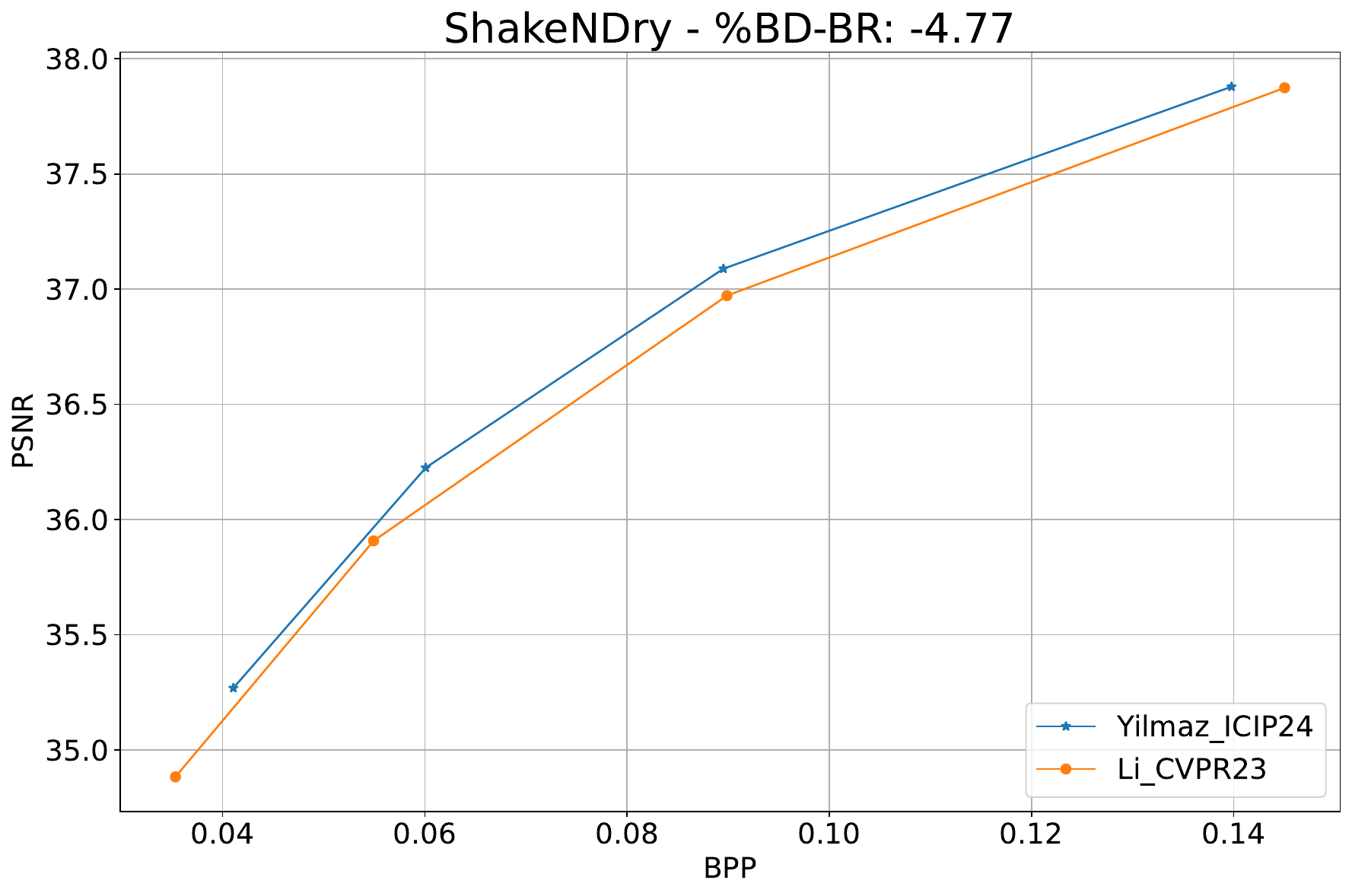}
}\hspace{-8pt}
\subfloat     {
\includegraphics[width=.248\linewidth]{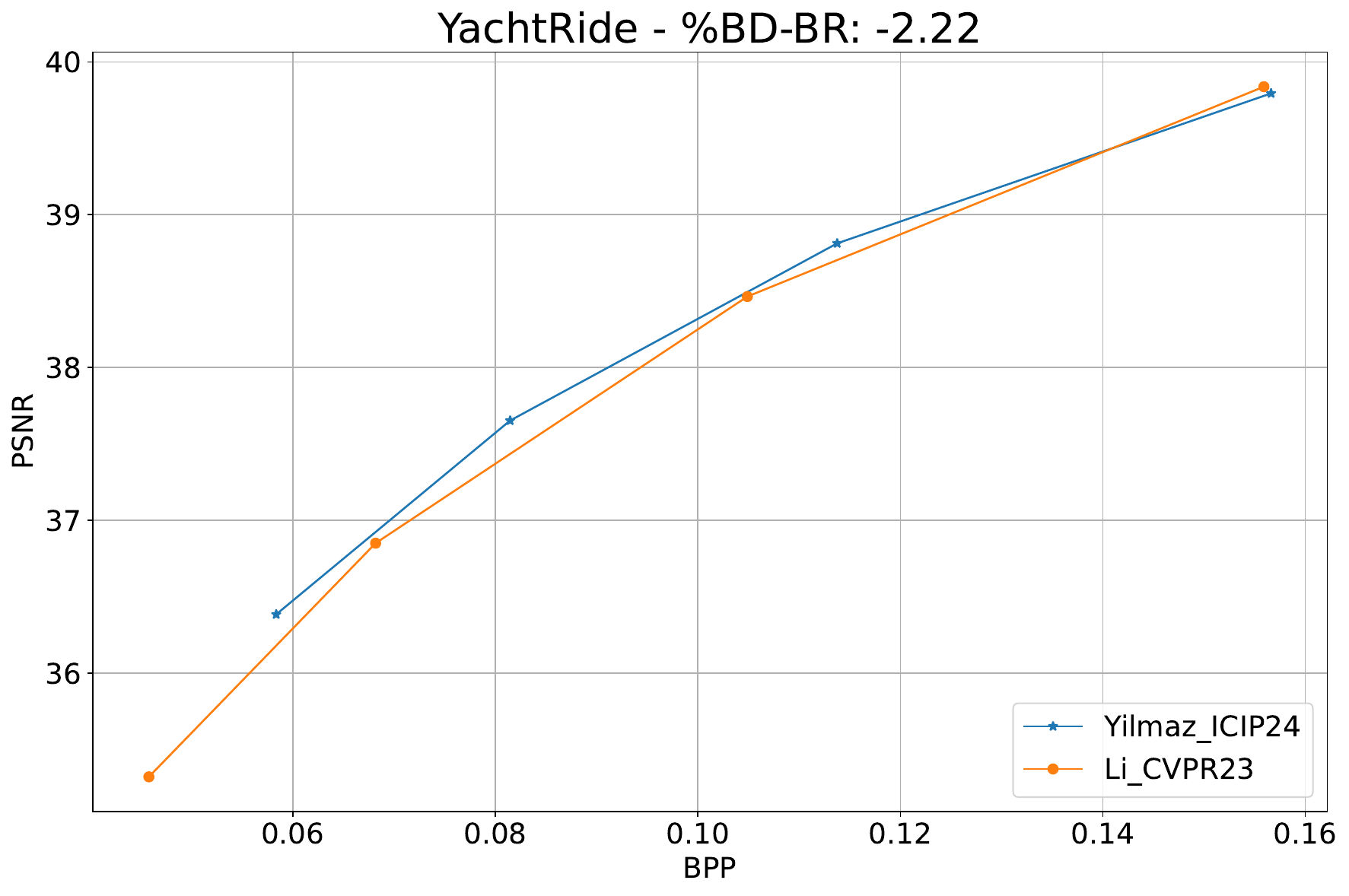}
} \hspace{-8pt}
\subfloat     {
\includegraphics[width=.248\linewidth]{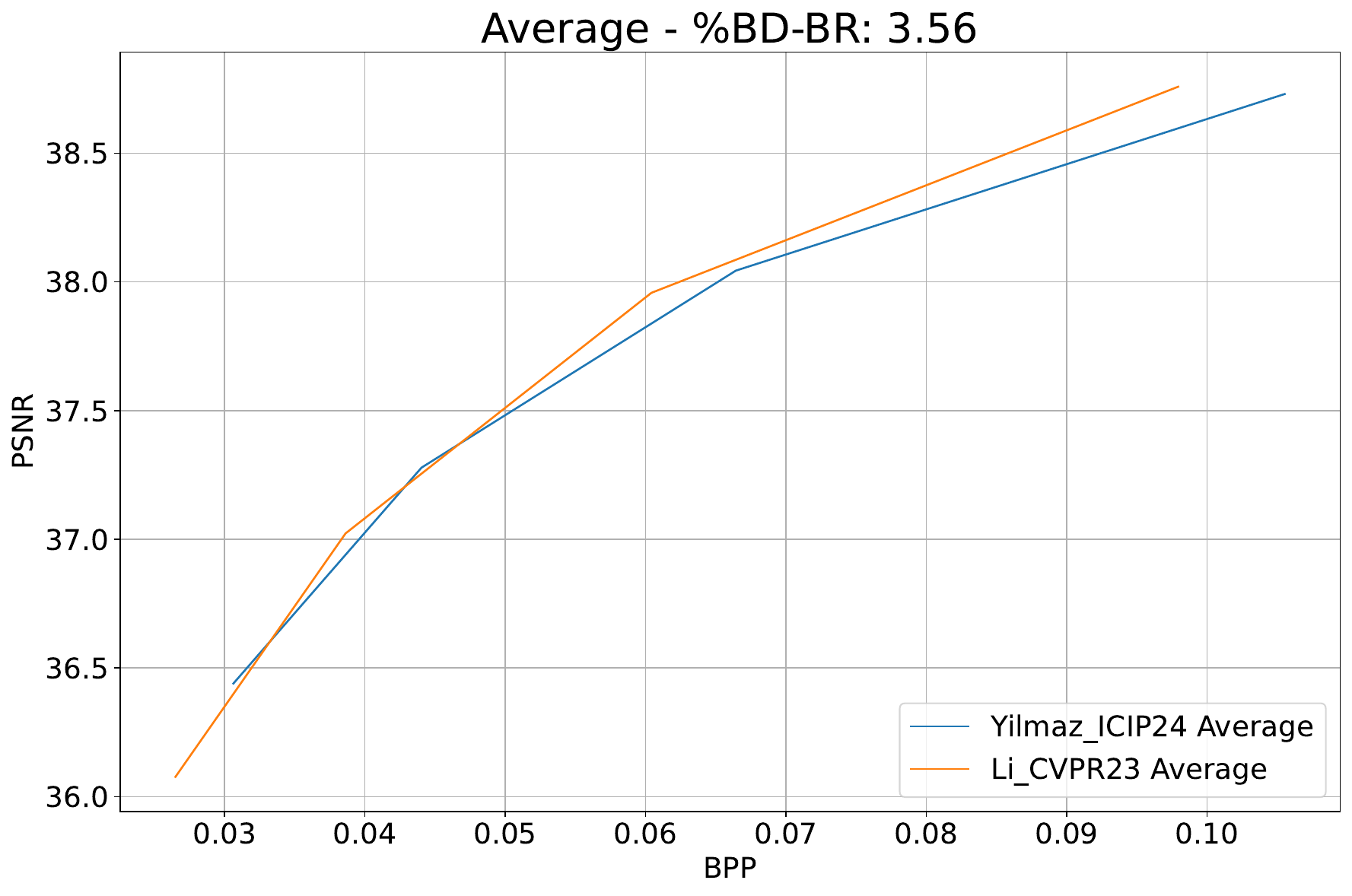}
} \vspace{-5pt}\\
\caption{Rate-Distortion curves for the seven (7) videos in the UVG\cite{uvg} dataset. The last graph is the average RD curve. Observe that \textit{ReadySetGo} is the only video where the average RD curve for Li{\_}CVPR2023 is on top of that of Yilmaz{\_}ICIP2024.}
\label{fig:uvg}
\end{figure*}

\begin{table*}[t]
\caption{BD-BR values for Yılmaz{\_}ICIP2024 in reference to Li{\_}CVPR2023. Observe that computing average of per-video BD-BR values vs. BD-BR from the average RD curve lead to conflicting comparative performances. } \vspace{-6pt}
\resizebox{\textwidth}{!}{ \begin{tabular}{|c|c|c|c|c|c|c|c|c|}
\hline
\textbf{Beauty} & \textbf{Bosphorus} & \textbf{Honeybee} & \textbf{Jockey} & \textbf{ReadySetGo} & \textbf{ShakeNDry} & \textbf{YachtRide} & \textbf{\begin{tabular}[c]{@{}c@{}}Average of\\ BD-BRs\end{tabular}} & \textbf{\begin{tabular}[c]{@{}c@{}}BD-BR on average\\  RD Curve\end{tabular}} \\ \hline
-5.43           & -25.16             & -23.19            & -5.27           & 16.80               & -4.77              & -2.22              & -7.03                                                                     & 3.56                                                                          \\ \hline
\end{tabular} } \vspace{-6pt}
\label{table:bdbr}
\end{table*}

\section{Experimental Evidence}
\label{sec:results}

\subsection{Experimental Setup}
To demonstrate how averaging RD curves can lead to misleading conclusions about codec performance, we analyze two recent learned video compression models: Li{\_}CVPR2023 (DCVC-DC)\cite{li2023neural}, which performs sequential coding, and learned bidirectional coding Yilmaz{\_}ICIP2024\cite{oursicip2024}. These models provide an excellent case study, as one model shows superior performance when evaluated properly on a per-sequence basis, yet appears inferior when assessed using averaged RD curves. We conducted experiments using the UVG dataset\cite{uvg}, which contains seven videos with diverse characteristics at 1920 $\times$ 1080 resolution. This dataset is particularly suitable for our analysis as it includes sequences with varying complexity, motion patterns, and operating ranges - factors that can significantly impact how averaging affects performance assessment.

\begin{figure}[t]
\centering
	\includegraphics[scale=0.245]{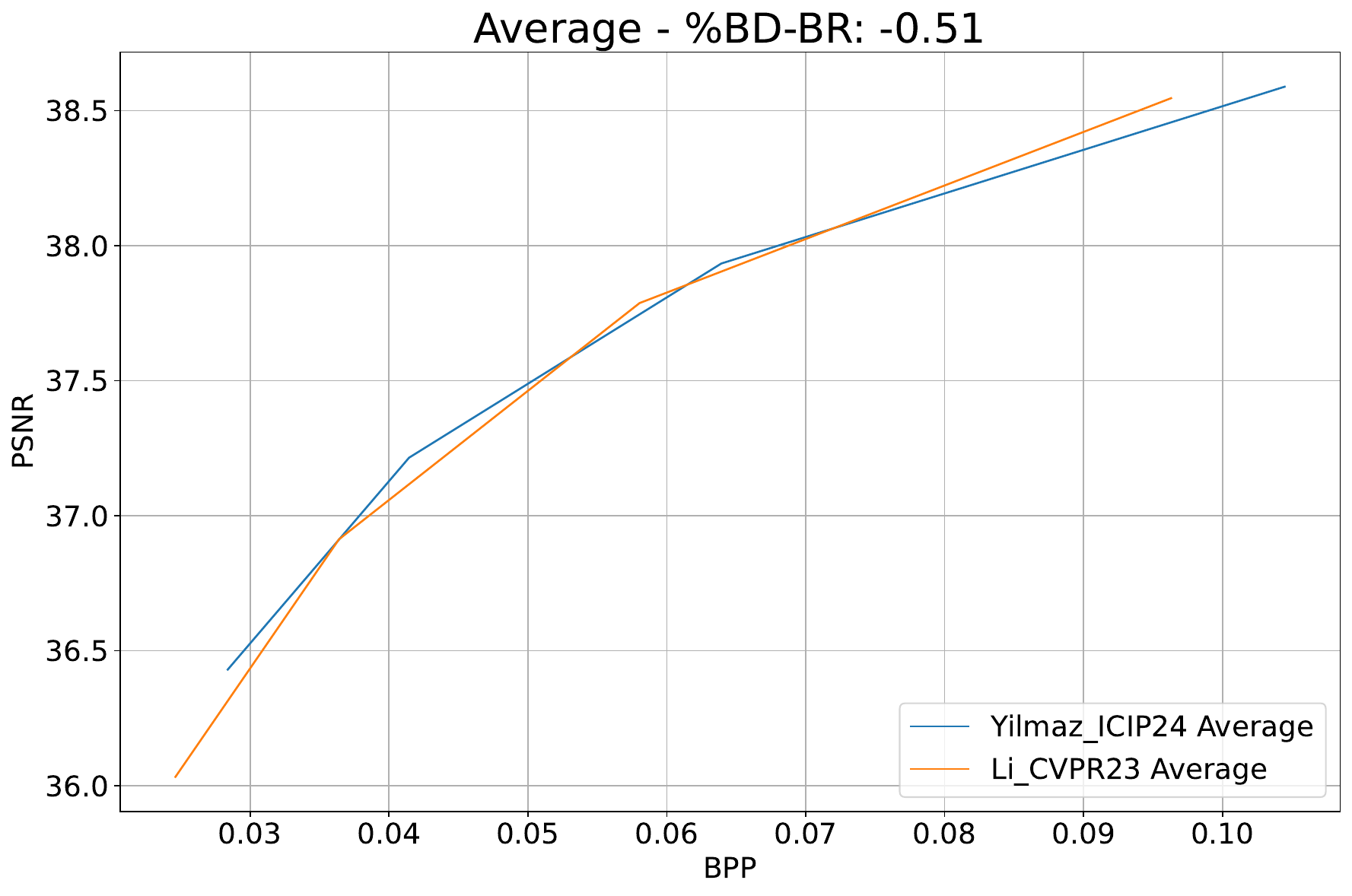} \vspace{-8pt} \\
\caption{Average RD curve for the UVG dataset if \textit{ReadySetGo} is excluded. The average RD curve for Li{\_}CVPR2023 is on top of that of Yilmaz{\_}ICIP2024, although we removed the only video where the curve for Li{\_}CVPR2023 is on top of that of Yilmaz{\_}ICIP2024.}
\label{fig:woready}
\end{figure}

\begin{figure}[t]
\centering
	\includegraphics[scale=0.245]{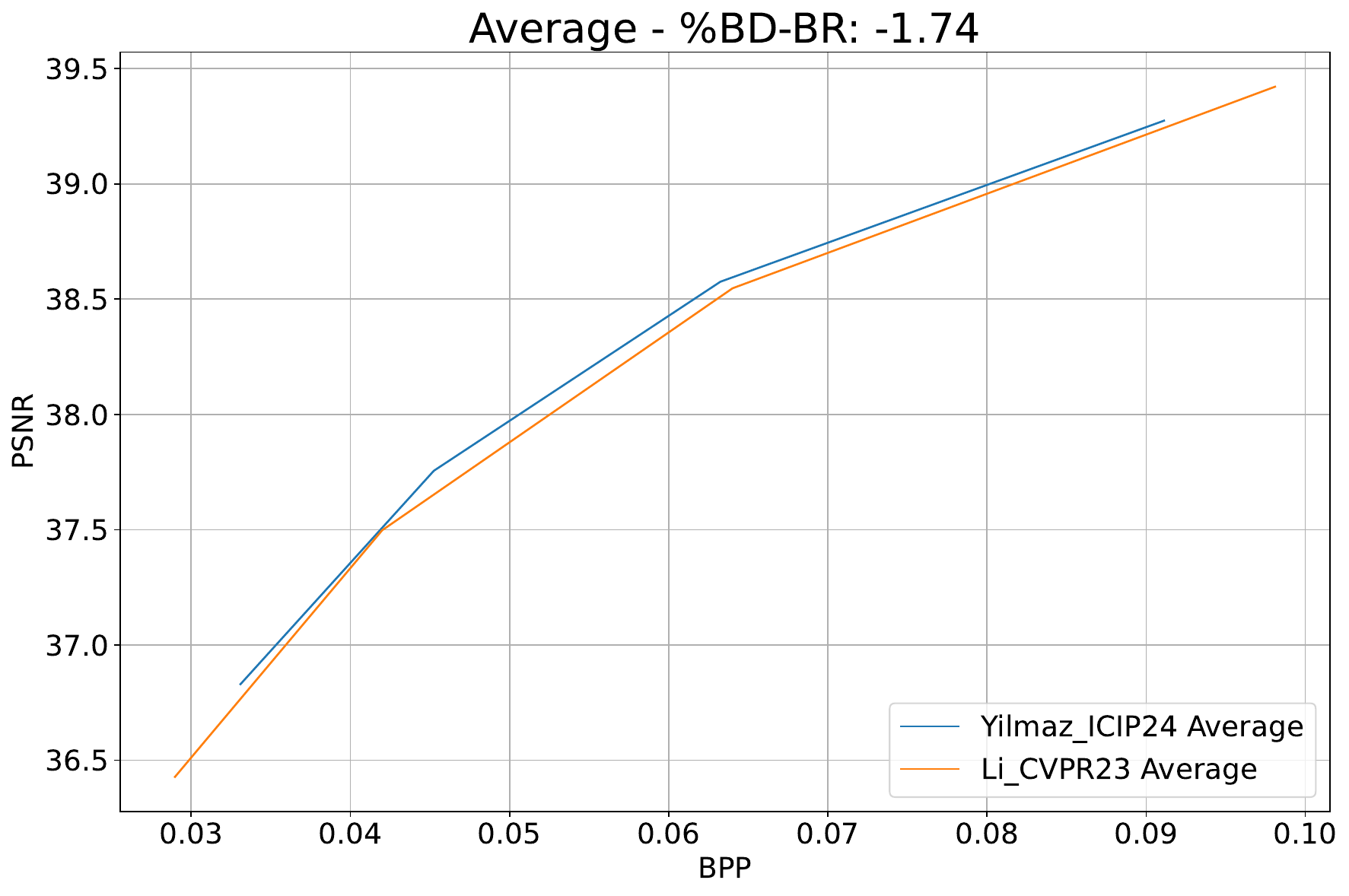} \vspace{-8pt} \\
\caption{Average RD curve for the UVG dataset if \textit{Beauty} is excluded. Excluding \textit{Beauty}, where Yilmaz{\_}ICIP2024 is superior, puts the RD curve for Yilmaz{\_}ICIP2024 over that of Li{\_}CVPR2023, which is inconsistent with other average RD curves. }
\label{fig:wobeauty}
\end{figure}

\vspace{-3pt}
\subsection{Analysis of RD Curve Averaging Effects}

Fig.~\ref{fig:uvg} presents per-sequence and averaged RD curves (PSNR vs. bits-per-pixel (bpp)) for both models. The per-sequence plots reveal that Yilmaz{\_}ICIP2024 outperforms Li{\_}CVPR2023 on all sequences except ReadySetGo. The performance differences vary significantly across sequences, with substantial gains on sequences like Bosphorus (-25.16\% BD-BR) and Honeybee (-23.19\% BD-BR), as shown in Table 1.

However, examining the averaged RD curve (last plot in Fig.~\ref{fig:uvg}) leads to a contradictory conclusion. Despite superior performance on most individual sequences, the averaged curve suggests Yilmaz{\_}ICIP2024 is 3.56\% inferior. This misleading result stems from two key factors:

1. Operating Range Differences: The Beauty sequence operates in a distinctly different quality range, with its maximum PSNR being lower than other sequences' minimum values. When averaged, this sequence disproportionately pulls  the overall curve down. 

2. Sequence Characteristics: Videos like Beauty, with complex motion patterns, saturate quickly in quality with limited PSNR improvements at high bitrates. In contrast, simpler sequences like Bosphorus and Honeybee achieve high PSNR values even at lower bitrates, creating an upward-shifting effect on the average curve.

\vspace{-3pt}
\subsection{Impact of Individual Sequences}

To further illustrate how averaging RD curves can mask true performance differences, we conducted two experiments:

First, we excluded ReadySetGo - the only sequence where Yilmaz{\_}ICIP2024 shows inferior performance. As shown in Fig.~\ref{fig:woready}, despite Yilmaz{\_}ICIP2024 being superior on all remaining sequences, the averaged RD curve still shows Li{\_}CVPR2023 outperforming at high bitrates - a clearly inconsistent result.

Second, we excluded the Beauty sequence (Fig.~\ref{fig:wobeauty}). Despite Yilmaz{\_}ICIP2024 showing superior performance on this sequence, its exclusion dramatically changes the average RD curve comparison. The model that appeared inferior when all sequences were included (3.56\% worse) now demonstrates superior performance (-1.74\% BD-BR).

These results highlight how averaging RD curves can lead to conclusions that directly contradict per-sequence performance metrics. The practice of averaging RD curves, while common in the learned video compression community, fails to capture the true relative strengths of different codecs and can lead to misleading assessments of codec performance.

\section{Conclusion}

This paper exposes a methodological issue in the evaluation of learned video compression models: the widespread~adopted practice of averaging RD curves across video sequences leads to potentially misleading performance assessments. Through both analytical analysis and real-world experiments, we demonstrate how this common practice can affect codec performance evaluation and even reverse comparative outcomes. Our results reveal that a codec showing superior performance on most individual sequences can paradoxically appear inferior when evaluated using average RD~curves.

The implications of our findings extend beyond just BD-rate computation. Our experimental results demonstrate how sequence-specific characteristics and varying operating ranges can disproportionately influence averaged RD curves. This influence becomes particularly problematic when different codecs operate in varying bitrate ranges - a common scenario in learned compression where models may adapt differently to content characteristics.

Most importantly, we demonstrate that averaging per-video performance metrics, rather than averaging RD curves, is a more reliable and fair assessment methodology. We emphasize that while this is already an established approach in the traditional video coding community, {\it there is no published literature that shows the pitfalls of computing a single BD rate from the average RD curve}, which is the de facto approach in the learned video compression community.

As the field of learned video compression continues to advance rapidly, ensuring fair and accurate performance assessment becomes increasingly crucial. We encourage researchers to carefully consider these findings and adopt per-sequence evaluation practices, as misleading performance metrics could significantly impact research directions and algorithmic choices in this rapidly evolving field.


\clearpage
\bibliography{references}

\begin{thebibliography}{10}
\providecommand{\url}[1]{#1}
\csname url@samestyle\endcsname
\providecommand{\newblock}{\relax}
\providecommand{\bibinfo}[2]{#2}
\providecommand{\BIBentrySTDinterwordspacing}{\spaceskip=0pt\relax}
\providecommand{\BIBentryALTinterwordstretchfactor}{4}
\providecommand{\BIBentryALTinterwordspacing}{\spaceskip=\fontdimen2\font plus
\BIBentryALTinterwordstretchfactor\fontdimen3\font minus \fontdimen4\font\relax}
\providecommand{\BIBforeignlanguage}[2]{{%
\expandafter\ifx\csname l@#1\endcsname\relax
\typeout{** WARNING: IEEEtran.bst: No hyphenation pattern has been}%
\typeout{** loaded for the language `#1'. Using the pattern for}%
\typeout{** the default language instead.}%
\else
\language=\csname l@#1\endcsname
\fi
#2}}
\providecommand{\BIBdecl}{\relax}
\BIBdecl

\bibitem{bdrate}
G.~Bjøntegaard, ``Improvements of the {BD-PSNR} model,'' \emph{Technical Report, VCEG-AI11, ITU-T SG16/Q6, Berlin, Germany}, 2008.

\bibitem{psnr_comp}
O.~Keleş, M.~A. Yilmaz, A.~M. Tekalp, C.~Korkmaz, and Z.~Doğan, ``On the computation of {PSNR} for a set of images or video,'' \emph{Picture Coding Symposium (PCS)}, pp. 1--5, 2021.

\bibitem{hevc}
G.~J. Sullivan, J.-R. Ohm, W.-J. Han, and T.~Wiegand, ``Overview of the high efficiency video coding (hevc) standard,'' \emph{IEEE Trans. on Circuits and Systems for Video Tech.}, vol.~22, no.~12, pp. 1649--1668, 2012.

\bibitem{vvc}
B.~Bross, Y.-K. Wang, Y.~Ye, S.~Liu, J.~Chen, G.~J. Sullivan, and J.-R. Ohm, ``Overview of the versatile video coding (vvc) standard and its applications,'' \emph{IEEE Trans. on Circuits and Systems for Video Tech.}, vol.~31, no.~10, pp. 3736--3764, 2021.

\bibitem{agustsson_scale}
E.~Agustsson, D.~Minnen, N.~Johnston, J.~Ballé, S.~J. Hwang, and G.~Toderici, ``Scale-space flow for end-to-end optimized video compression,'' in \emph{IEEE/CVF Conf. Comp. Vis. Patt. Recog. (CVPR)}, 2020, pp. 8500--8509.

\bibitem{rlvc}
R.~Yang, F.~Mentzer, L.~Van~Gool, and R.~Timofte, ``Learning for video compression with recurrent auto-encoder and recurrent probability model,'' \emph{IEEE Jour. of Selected Topics in Signal Processing}, vol.~15, no.~2, pp. 388--401, 2021.

\bibitem{elfvc}
O.~Rippel, A.~G. Anderson, K.~Tatwawadi, S.~Nair, C.~Lytle, and L.~Bourdev, ``Elf-vc: Efficient learned flexible-rate video coding,'' \emph{IEEE/CVF Int. Conf. on Computer Vision (ICCV)}, 2021.

\bibitem{FVC2022}
Z.~Hu, D.~Xu, G.~Lu, W.~Jiang, W.~Wang, and S.~Liu, ``Fvc: An end-to-end framework towards deep video compression in feature space,'' \emph{IEEE Trans. on Pattern Analysis and Machine Intelligence}, pp. 1--17, 2022.

\bibitem{ladune2021conditional}
T.~Ladune, P.~Philippe, W.~Hamidouche, L.~Zhang, and O.~D{\'e}forges, ``Conditional coding for flexible learned video compression,'' in \emph{Neural Compression: From Info. Theory to Applications, ICLR Workshop}, 2021.

\bibitem{li2021deep}
J.~Li, B.~Li, and Y.~Lu, ``Deep contextual video compression,'' \emph{Advances in Neural Information Processing Systems}, vol.~34, 2021.

\bibitem{li2022hybrid}
------, ``Hybrid spatial-temporal entropy modelling for neural video compression,'' in \emph{Proc. of the~30th ACM Int. Conf. on Multimedia}, 2022.

\bibitem{sheng2022temporal}
X.~Sheng, J.~Li, B.~Li, L.~Li, D.~Liu, and Y.~Lu, ``Temporal context mining for learned video compression,'' \emph{IEEE Transactions on Multimedia}, 2022.

\bibitem{li2023neural}
J.~Li, B.~Li, and Y.~Lu, ``Neural video compression with diverse contexts,'' in \emph{{IEEE/CVF} Conf. on Comp. Vis. and Patt. Recog. {CVPR}, Vancouver, Canada, 2023}, 2023.

\bibitem{canfvcpp}
\BIBentryALTinterwordspacing
P.-Y. Chen and W.-H. Peng, ``Canf-vc++: Enhancing conditional augmented normalizing flows for video compression with advanced techniques,'' 2023. [Online]. Available: \url{https://arxiv.org/abs/2309.05382}
\BIBentrySTDinterwordspacing

\bibitem{hlvc}
R.~Yang, F.~Mentzer, L.~Van~Gool, and R.~Timofte, ``Learning for video compression with hierarchical quality and recurrent enhancement,'' in \emph{IEEE/CVF Conf. on Computer Vision and Patt. Recog. (CVPR)}, 2020.

\bibitem{hinerv}
H.~M. Kwan, G.~Gao, F.~Zhang, A.~Gower, and D.~Bull, ``Hinerv: Video compression with hierarchical encoding based neural representation,'' 2023.

\bibitem{chen2021nerv}
H.~Chen, B.~He, H.~Wang, Y.~Ren, S.-N. Lim, and A.~Shrivastava, ``Ne{RV}: Neural representations for videos,'' in \emph{NeurIPS}, 2021.

\bibitem{ours_icip20}
M.~A. Yilmaz and A.~M. Tekalp, ``End-to-end rate-distortion optimization for bi-directional learned video compression,'' in \emph{IEEE Int. Conf. on Image Processing (ICIP)}, 2020, pp. 1311--1315.

\bibitem{lhbdc}
M.~A. Yılmaz and A.~M. Tekalp, ``End-to-end rate-distortion optimized learned hierarchical bi-directional video compression,'' \emph{IEEE Transactions on Image Processing}, vol.~31, pp. 974--983, 2022.

\bibitem{flexrate}
E.~Çetin, M.~A. Yılmaz, and A.~M. Tekalp, ``Flexible-rate learned hierarchical bi-directional video compression with motion refinement and frame-level bit allocation,'' in \emph{IEEE Int. Conf. on Image Processing (ICIP)}, 2022, pp. 1206--1210.

\bibitem{ours_icip23}
M.~A. Yılmaz, O.~Ugur~Ulas, and A.~M. Tekalp, ``Multi-scale deformable alignment and content-adaptive inference for flexible-rate bi-directional video compression,'' in \emph{IEEE Int. Conf. on Image Processing (ICIP)}, 2023, pp. 2475--2479.

\bibitem{pytorchvideocompression}
Z.~Hu, ``Pytorch video compression benchmark,'' \url{https://github.com/ZhihaoHu/PyTorchVideoCompression/tree/master/Benchmark}, 2022, accessed: Dec. 2024.

\bibitem{uvg}
A.~Mercat, M.~Viitanen, and J.~Vanne, ``{UVG} dataset: 50/120fps 4k sequences for video codec analysis and development,'' in \emph{ACM Multimedia Systems Conference}, ser. MMSys '20, 2020, p. 297–302.

\bibitem{oursicip2024}
M.~A. Yilmaz, O.~U. Ulas, A.~Bilican, and A.~M. Tekalp, ``Motion-adaptive inference for flexible learned b-frame compression,'' in \emph{IEEE Int. Conf. Image Processing (ICIP)}, 2024.

\end{thebibliography}
\bibliographystyle{IEEEtran}

\end{document}